\documentclass[aps,prb,twocolumn,showpacs,groupedaddress]{revtex4-1}  
\usepackage{graphicx}  
\usepackage{dcolumn}   
\usepackage{bm}        
\usepackage{amssymb}   
\usepackage{amsmath}
\usepackage{latexsym}
\usepackage{amsfonts}
\usepackage{amstext}
\usepackage{graphicx}
\usepackage{graphics}
\graphicspath{{./img/}}

\begin{document}



\title{Control of the transmission phase in an asymmetric four-terminal Aharonov-Bohm interferometer}
\date{\today}

\author{
Sven S. Buchholz, Saskia F. Fischer}\email[electronic address: ]{saskia.fischer@rub.de} \author{Ulrich Kunze\\
\small Lehrstuhl f\"ur Werkstoffe und Nanoelektronik, Ruhr-Universit\"at Bochum, 44780 Bochum, Germany\\
\normalsize Matthew Bell\\
\small Electrical Engineering Department, University at Buffalo, New York 14260, USA\\
\normalsize Dirk Reuter, Andreas D. Wieck\\
\small Lehrstuhl f\"ur Angewandte Festk\"orperphysik, Ruhr-Universit\"at Bochum, 44780 Bochum, Germany}

\begin{abstract}
Phase sensitivity and thermal dephasing in coherent electron transport in quasi one-dimensional (1D) waveguide rings of an asymmetric four-terminal geometry are studied by magnetotransport measurements. We demonstrate the electrostatic control of the phase in Aharonov-Bohm (AB) resistance oscillations and investigate the impact of the measurement
circuitry on decoherence. Phase rigidity is broken due to the ring geometry: Orthogonal waveguide cross-junctions and 1D leads minimize reflections and resonances between leads
allowing for a continuous electron transmission phase shift. The measurement circuitry influences dephasing: Thermal averaging dominates in the non-local measurement configuration while additional influence of potential fluctuations becomes relevant in the local configuration.
\end{abstract}

\maketitle

\section{Introduction}

Novel quantum electronic devices are of prime research interest 
with regard to the control and manipulation of coherence and phase of the electron wave function in a solid state environment. Electron interference in paths enclosing a magnetic flux, known as the Aharonov-Bohm (AB) effect \cite{ahar59}, enables to monitor both coherence and phase properties in ballistic quantum rings by measuring magnetoresistance oscillations. Utilizing the AB effect, coherent transmission through a quantum dot (QD) has been demonstrated \cite{yac95}, and the transmission phase of QDs has been probed recently \cite{schu97,koba0204,sigr04}. Electron correlation \cite{ji00} and spin effects \cite{koen02}, in particular in combination with spin filters \cite{koen10} are invoking considerable research interest. For this purpose the reliable experimental detection and control of the transmission phase is required, which remains a challenge.

In order to determine the transmission phase evolution from resistance oscillations it is necessary to overcome the restriction of the AB phase to zero or $\pi$ at zero magnetic field (phase rigidity) \cite{yaco96}. In the linear transport regime phase rigidity can only be broken if the following conditions are met: (a) The scattering matrix unitarity of the current carrying leads needs to be broken by an open multi-terminal geometry \cite{butt88,enti02}, (b) geometrical device symmetries leading to symmetries of transmission coefficients need to be broken \cite{krei10}, and (c) electron reflections at current and voltage leads need to be reduced to a minimum. Reflections and resonances can lead to a loss of the actual transmission phase of the interference paths \cite{datt89,enti02} and yield abrupt $\pi$ phase jumps \cite{krei10}. These conditions demand a multi-terminal, asymmetric AB-interferometer geometry with minimal internal reflections.

To date, the temperature dependence of dephasing in quantum wire rings \cite{aiha91,cass00,hans01,koba02,yama09,lin10} calls for an investigation with respect to the measurement circuitry \cite{koba02,lin10}, ring geometry \cite{seel03}, as well as the effect of gates \cite{seel01}. At low temperatures, coherence in ballistic conductors is limited mainly by electron-electron scattering \cite{alts82} and thermal averaging \cite{wash85,cass00,hans01}. Additionally, decoherence due to charge fluctuations has been suggested \cite{seel01} 
and discussed to evoke a measurement-configuration-dependence of dephasing with temperature \cite{seel03}. Recently, a strong configuration-dependence has been observed in a symmetric AB ring with finger-top-gates \cite{koba02}.
 
In this work, we report on the electrostatic control of the transmission phase and on temperature dephasing in \textit{strongly asymmetric} quantum wire rings with quantum wire current leads and voltage probes and \textit{global top-gates}. 
This realization allows for an effectively four-terminal measurement which breaks phase rigidity and -- in terms of dephasing -- is sensitive to the measurement circuitry.

\section{Device geometry and measurement details}

\begin{figure}[t]
\begin{center}
\includegraphics[width=0.85\columnwidth]{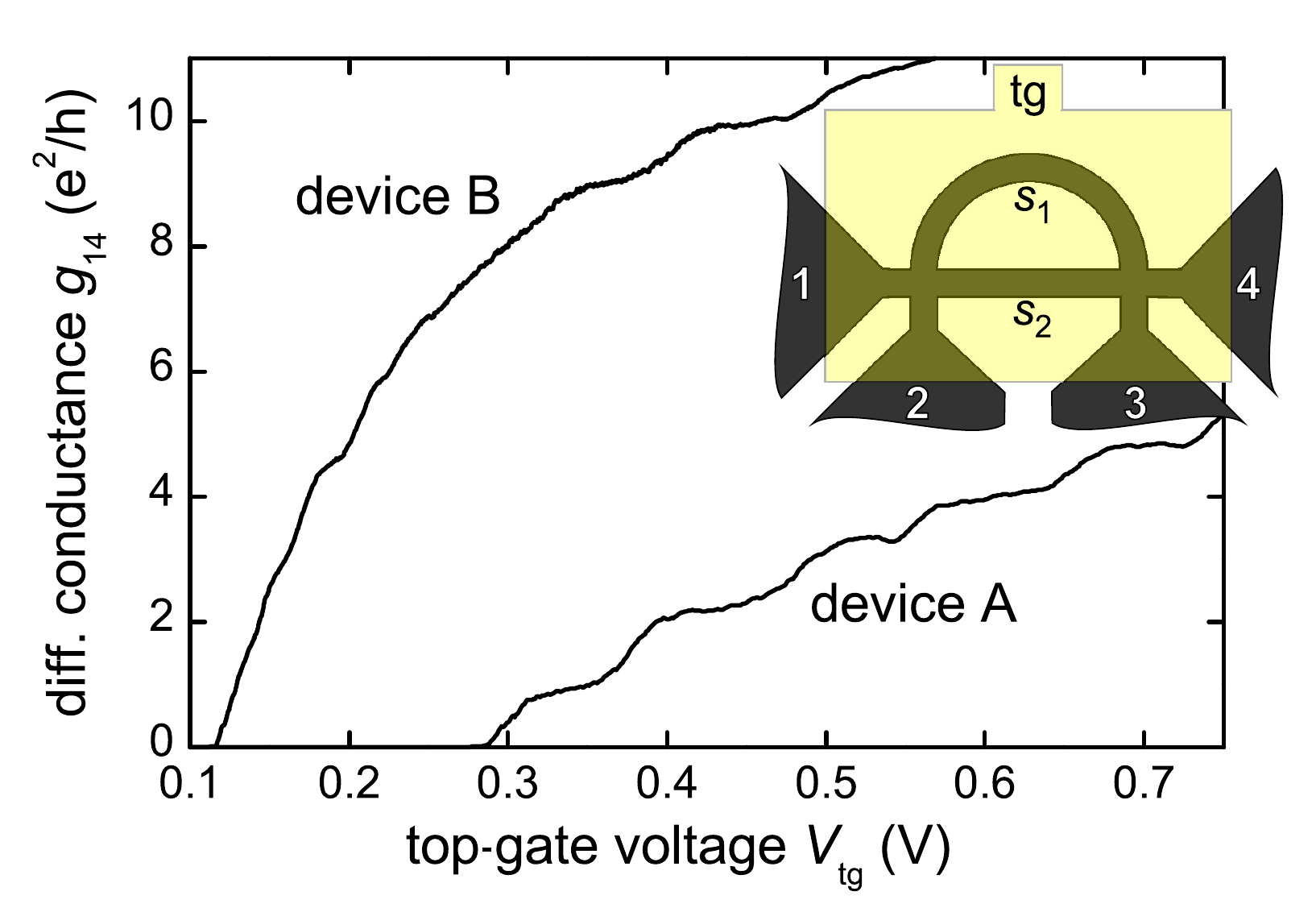}
\caption{(Color online) Two-terminal differential conductance $g_{14}$ measured at $T=4.2$~K and zero magnetic field as a function of the top-gate voltage $V_\mathrm{tg}$. 1D transport characteristics are manifested in conductance quantization. The inset depicts a schematic of a waveguide quantum ring with adjacent quantum wire leads which link to 2D reservoirs. The yellow rectangle represents the global top-gate (tg) electrode.}
\label{gdVtg}
\end{center}
\end{figure}

Two similar AB rings were investigated. The quantum rings comprise one straight and one half-circle 1D waveguide which leads to geometrically asymmetric electron paths in the AB experiment (see inset of Fig.~\ref{gdVtg}). The 1D waveguide rings are connected to 2D reservoirs via short 1D quantum wire (QWR) leads, and all 1D structures are of the same geometric width $w$ and etching depth. This guarantees a collimated and single-mode controlled electron injection into the ring. The AB effect was observed in these rings as $h/e$-magnetoresistance oscillations in two- and four-terminal measurements at temperatures as high as $T=1.5$~K \cite{buch09a,buch09b}.

The waveguide quantum rings were prepared from a modulation doped AlGaAs/GaAs heterostructure grown by molecular-beam epitaxy. A two-dimensional electron gas (2DEG) is situated at the heterojunction 55 nm below the surface and has an electron density of $3.1\times10^{11}$~cm$^{-2}$ and a mobility of $1\times10^6$~cm$^2/\mathrm{Vs}$ measured in the dark at $T=4.2$~K. This corresponds to an elastic electron mean free path of $l_{\mathrm{e}}\approx9.5$~$\mu$m. The nanoscale devices were fabricated using electron-beam lithography and wet-chemical etching with an etch depth of approximately 40 nm. The geometric width of the etched waveguides amounts to $w\approx150$~nm (device A) and 250~nm (device B), the distances between the intersection centers (cross-junctions) of the waveguides are $s_1\approx3.3$~$\mu$m along the bent and $s_2\approx2$~$\mu$m along the straight waveguides, encircling an area of $A\approx1.7$~$\mu$m$^2$. The total lengths of the straight and the bent waveguides are 2.85 and 4.15~$\mu$m, respectively, where each QWR lead from the 2D reservoirs to the waveguide ring is approximately 300~nm long and of the same width $w$ as the waveguides. In order to control the electron densities in each device a Ti/Au top-gate electrode covers all waveguide structures, the leads as well as parts of the 2D reservoirs. The inset of Fig.~\ref{gdVtg} shows a schematic of the devices, a scanning electron micrograph of device B can be found elsewhere~\cite{buch09a}.
Devices A and B solely differ in the waveguide widths $w$ which determines the 1D confinement strength.

Transport experiments were performed in the mixing chamber of a dilution refrigerator at a base temperature of $T_\mathrm{base}=23$~mK and an electron temperature of approximately 100~mK. Measurements at $T=4.2$~K were carried out in a liquid helium dewar. Current-voltage (I-V) characteristics were recorded in the quasi-linear transport regime. In two-terminal configurations, we measured the differential conductance $(dI/dV)^{-1}$ using standard lock-in technique with an ac excitation voltage of $10$~$\mu$V at 133~Hz. Four-terminal measurements were performed at excitation currents of $5$~nA to 10~nA at 77 or 133~Hz. The magnetic field was swept quasi-statically in steps of 50~$\mu$T or less.

Two four-probe resistance measurement configurations $R_{ij,kl}=(V_k-V_l)/I_{ij}=V_{kl}/I_{ij}$ were set up: (a) In the local configuration, the voltage probes are placed along the current path, e.g. $R_{14,23}$. (b) In the non-local configuration, the voltage probes are spatially detached from the current path, e.g. $R_{34,21}$. Thus, the non-local voltage-current measurement provides the transfer characteristic.

\section{Experimental results}

\subsection{Ballistic transport}

A characteristic feature of 1D ballistic transport is conductance quantization \cite{vanW88,whar88} in units of $g_0=2e^2/h$ in GaAs electron waveguides \cite{lian99,knop05}. In parallel waveguides or quantum rings, the quantization will appear in units smaller than $g_0$
, which can be approximated by a classical resistor network. A quantum ring with two leads and two arms will ideally show a conductance quantization in units of $(1/g_0+1/(2g_0)+1/g_0)^{-1}=0.8e^2/h$.

Fig.~\ref{gdVtg} depicts the two-terminal differential conductance $g_\mathrm{14}$ as a function of the top-gate voltage $V_\mathrm{tg}$ measured between probes 1 and 4 for both devices A and B at $T=4.2$~K. The measurements feature conductance quantization in units of $1e^2/h$ to $1.3e^2/h$ which are higher than expected from the simple resistor model. A simple series and parallel connection of resistors does not fully apply to our 1D waveguide network in which the orthogonal cross-junctions partially maintain the 1D transport character. As expected, the conductance characteristic of device A has a higher threshold voltage and more significant plateaux than that of device B because it exhibits a stronger 1D confinement. From the conductance measurements at 4.2~K we can estimate the number of populated subbands in the waveguides at lower temperatures. In the gate voltage range of 0.6 to 0.73~V about 3 to 6 and 8 to 12 subbands were populated in the case of device A and B, respectively.

At temperatures below 1~K conductance fluctuations are superimposed on the characteristic measurements and finally wash out the conductance steps at the base temperature. These fluctuations with $V_\mathrm{tg}$ are attributed to universal conductance fluctuations and resonances in the waveguides as typical for complex mesoscopic systems with large phase coherence lengths \cite{hans01,krei10}.

\subsection{Electrostatic AB phase shift}

\begin{figure}[t]
\begin{center}
\includegraphics[width=1\columnwidth]{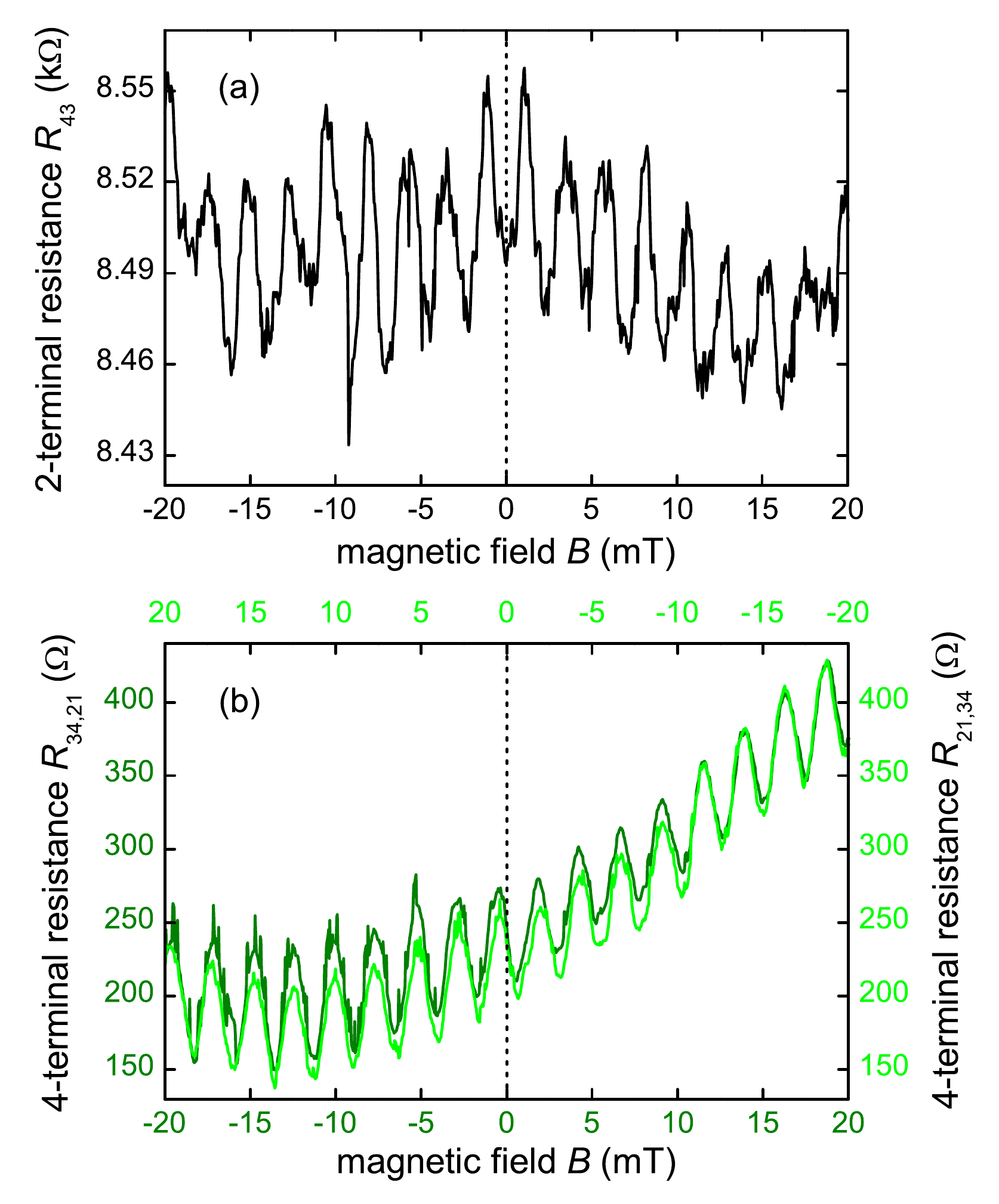}
\caption{(Color online) (a) Two-terminal measurement (raw data) of the magnetoresistance $R_{43}(B)$ at $T_\mathrm{base}=23$~mK. The symmetry around $B=0$ leads to the pinning of the AB phase to 0 or $\pi$ at $B=0$ (phase rigidity). (b) Non-local four-terminal measurements (raw data) of the magnetoresistance at $T_\mathrm{base}=23$~mK. The dark green curve shows $R_{34,21}$ (left axis) with increasing magnetic field (lower axis), whereas the light green curve depicts $R_{21,34}$ (right axis) with decreasing magnetic field (upper axis). Both resistance traces are almost identical in magnitude and phase. The absence of symmetry around $B=0$ indicates broken phase rigidity.}
\label{phaserig}
\end{center}
\end{figure}

Magnetotransport measurements at fixed $V_\mathrm{tg}$ visualize interference as $h/e$ resistance oscillations. At $T_\mathrm{base}=23$~mK these oscillations are well resolved with visibilities $v=(R_\mathrm{max}-R_\mathrm{min})/(R_\mathrm{max}+R_\mathrm{min})$ (peak-to-peak) of up to 0.6\% in two-terminal and 30\% in non-local four-terminal measurements. Fig.~\ref{phaserig} (a) shows the raw data of a typical two-terminal magnetoresistance measurement. The resistance oscillates in the magnetic field on an aperiodic background, and it shows a symmetry around $B=0$, concerning both the oscillatory part and the background resistance as expected due to time-reversal invariance (Onsager-Casimir relation \cite{onsa31}) and current conservation in a closed, i.e. unitary, interferometer. The transmission coefficients for an electron wave to propagate from terminal $i$ to terminal $j$ and vice versa are given by $T_{ij}(B)=T_{ji}(-B)$ \cite{butt86,butt88}. In a two-terminal conductor the requirement of current conservation yields the symmetry of the transmission coefficients $T_{ij}(B)=T_{ij}(-B)$ and hence $R_{ij}(B)=R_{ij}(-B)$. In consequence the phase of the AB resistance oscillation is pinned to zero or $\pi$ at zero magnetic field and can not evolve continuously but jumps abruptly by $\pi$ as it has been observed experimentally \cite{ped00,stra09}.

In a four-terminal configuration, time reversal invariance leads to the relation $R_{kl,mn}(B)=R_{mn,kl}(-B)$, and current conservation with respect to the measurement leads is no longer given, i.e. unitarity is broken. Thus, a symmetry of the four-terminal resistance $R_{kl,mn}$ does not generally emerge. Fig.~\ref{phaserig} (b) shows four-probe magnetoresistance measurements in the non-local configuration. The transfer-resistance $R_{34,21}$ is plotted from negative to positive magnetic field, whereas the measurement with interchanged voltage and current leads $R_{21,34}$ is plotted from positive to negative magnetic field. In contrast to the two-probe measurement, $R_{34,21}$ and $R_{21,34}$ are asymmetric in the magnetic field. Fig.~\ref{phaserig} (b) illustrates that $R_{34,21}(B)$ and $R_{21,34}(-B)$ measurements resemble each other and that the phase situation in both measurements is identical. The observation of $R_{kl,mn}(B)=R_{mn,kl}(-B)$ indicates that unitarity and phase rigidity of the waveguide quantum ring are broken in the non-local measurement configuration.

Further, we measured the magnetoresistance as a function of the gate voltage to detect the electrostatic part of the AB-effect \cite{datt89,wash87,deVeg89,ford90}. Due to the ring's geometrical asymmetry, a change of $k_\mathrm{F}$ evokes an AB oscillation phase shift: At a fixed magnetic field and neglecting constant phases $\varphi_i$, the interference of partial electron waves $\Psi_\mathrm{s_1}=|\Psi_\mathrm{s_1}|\cdot e^{\mathrm{i}k_{\mathrm{F}}s_1}$ propagating along the waveguide $s_1$ and $\Psi_\mathrm{s_2}=|\Psi_\mathrm{s_2}|\cdot e^{\mathrm{i}k_{\mathrm{F}}s_2}$ propagating along $s_2$ is $|\Psi_\mathrm{s1}+\Psi_\mathrm{s2}|^2=|\Psi_\mathrm{s1}|^2+|\Psi_\mathrm{s2}|^2+2|\Psi_\mathrm{s1}||\Psi_\mathrm{s2}|\mathrm{cos}[k_\mathrm{F}\Delta s]$, where
$\Delta s=s_1-s_2$ is the geometrical difference between the paths. We assume that the wave vector $k_\mathrm{F}$ in both paths is the same. Since $k_\mathrm{F}\Delta s$ is non-zero, and $k_\mathrm{F}$ depends on the gate voltage $k_\mathrm{F}(V_\mathrm{tg})$, an electrostatic control of the phase of AB resistance oscillations is provided by $V_\mathrm{tg}$.

In order to roughly estimate the electrostatically induced phase shift, we approximate $k_\mathrm{F}$ in a single electron 2D-model by first order as a linear function of $V_\mathrm{tg}$ around $V_\mathrm{tg}=0.7$~V for device B: $\Delta k_\mathrm{F}(\Delta V_\mathrm{tg})\Delta s\approx180 \mathrm{V}^{-1} \Delta V_\mathrm{tg}$, where the slope has been determined from Shubnikov-de Haas measurements of the electron densities via $R_{xx}=R_{14,23}$. From this approach, we expect $\Delta V_\mathrm{tg}\approx0.35$~V for a $2\pi$-shift of the AB oscillation phase.

Due to universal conductance fluctuations it is not possible to reveal the electrostatic AB effect directly as an oscillatory resistance modulation $R_{\mathrm{AB}}(V_\mathrm{tg})|_{B=\mathrm{const}}\propto \mathrm{cos}[k_\mathrm{F}(V_\mathrm{tg})\Delta s]$ by sweeping the gate voltage at a fixed magnetic field. The AB oscillations are significantly smaller than irregular $V_\mathrm{tg}$-dependent resistance fluctuations. However, magnetoresistance measurements at different gate voltages $R_{\mathrm{AB}}(B)|_{V_\mathrm{tg}=\mathrm{const}}$ can reveal an electrostatic phase shift since the $V_\mathrm{tg}$-dependent background resistance is irrelevant for the phase detection and can be subtracted.
\begin{figure*}[t]
\begin{center}
\includegraphics[width=1.8\columnwidth]{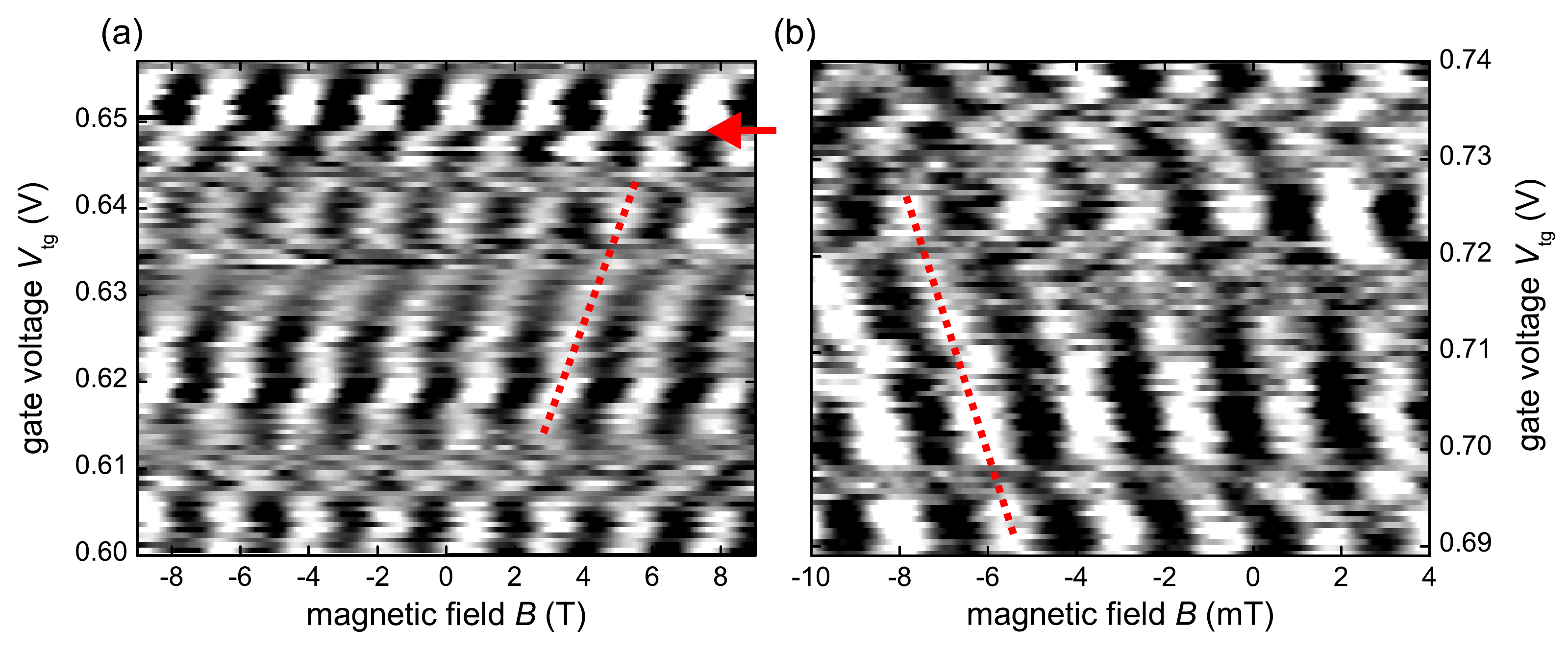}
\caption{(Color online) Grey scale plots of the oscillatory parts of the non-local magnetoresistances (a) $R_{21,34}$ of device A ($w=150$~nm) and (b) $R_{43,12}$ of device B ($w=250$~nm) versus top-gate voltage $V_{\mathrm{tg}}$ and magnetic field $B$. Magnetoresistance traces were recorded for succeeding gate voltages in steps of $\Delta V_{\mathrm{tg}}=0.6$~mV at $T_\mathrm{base}=23$~mK. The red dotted lines indicate an overall continuous phase shift, the red arrow indicates a typical abrupt phase jump by $\pi$.}
\label{phaseshift}
\end{center}
\end{figure*}

Fig.~\ref{phaseshift} depicts grey scale plots of the non-local AB resistance versus gate voltage and magnetic field. Magnetoresistance measurements were recorded for successive gate voltages at $T_\mathrm{base}=23$~mK, and the background resistance was subtracted. The AB grey scale plot of device A ($w=150$~nm) in shown in Fig.~\ref{phaseshift} (a), whereas Fig.~\ref{phaseshift} (b) shows the results of device B ($w=250$~nm). In both figures an overall phase shift is visible as indicated by the red dotted lines. In Fig.~\ref{phaseshift} (a) successive AB maxima of $R_{21,34}$ shift to higher magnetic fields with increasing gate voltage, i.e. the phase increases. In contrast, the measurement of $R_{43,12}$ of device B (Fig.~\ref{phaseshift} (b)) shows a decreasing phase with increasing gate voltage which finds a direct explanation in the different measurement configuration. In the non-local measurement of $R_{21,34}$ electrons are injected at the left side of the ring and propagate to the right, and vice versa for $R_{43,12}$. From Fig.~\ref{phaseshift} (b) we can extract that a voltage change of $\Delta V_\mathrm{tg}\approx0.36$~V is required for a $2\pi$-phase shift, which is in good agreement with our estimate of 0.35~V.
   
Next to the phase shift, further observations are remarkable: (a) regions of reduced AB amplitudes (e.g. Fig.~\ref{phaseshift} (a) around $V_\mathrm{tg}=0.63$~V), (b) abrupt phase jumps (e.g. Fig.~\ref{phaseshift} (a) around $V_\mathrm{tg}=0.649$~V as indicated by the red arrow) and (c) sometimes higher harmonics (h/2e oscillations). 
These irregular features are superimposed on the electrostatic AB oscillation phase shift, and we suggest electron wave scattering and reflections in the waveguide cross-junctions as possible causes.

A theoretical approach of Kreisbeck and Kramer \textit{et al.} \cite{kram08,krei10} includes scattering and reflections at realistically designed waveguide cross-junctions in a corresponding waveguide quantum ring. The calculation results qualitatively reproduce our experimental findings: In the non-local four-terminal setup the AB oscillation phase shifts continuously in certain energy ranges which are interrupted by irregular abrupt $\pi$ phase jumps, regions of reduced visibility and higher harmonic oscillations. These irregular features are mainly attributed to Fermi-energy-dependent scattering, reflections and resonances at the waveguide cross-junctions \cite{krei10}.

In recent four-terminal ballistic quasi-1D rings \cite{deVeg89,ford90,cern97,kraf01} a continuous AB oscillation phase shift was not observed for three reasons: 
Firstly, geometrical symmetries impose symmetries on the transmission coefficients which force the four-terminal resistance to be an even function in the magnetic field \cite{krei10}, secondly, the electron phase-coherence length may be too small the ring extensions \cite{butt86,ben86}, and thirdly, reflections at the leads and resonances between the leads impose symmetry under magnetic field reversal because the transmission phases picked up along the waveguides cancel out over the reflections \cite{enti02}.

We conclude, that in our devices, the ring dimensions and the QWR leads which connect the 2D reservoirs with the waveguide ring play a decisive role to break phase rigidity: The current and voltage probes are in closest vicinity to the ring, and each lead connects the ring separately. In addition the 2D-1D mode-matching resistances between the 2D reservoirs and the QWR leads occur outside of the AB ring and the voltage/current leads. Orthogonal crossings of the waveguides at the leads minimize scattering and reflections.

\subsection{Measurement-configuration-dependent dephasing}

\begin{figure}[t]
\begin{center}
\includegraphics[width=0.9\columnwidth]{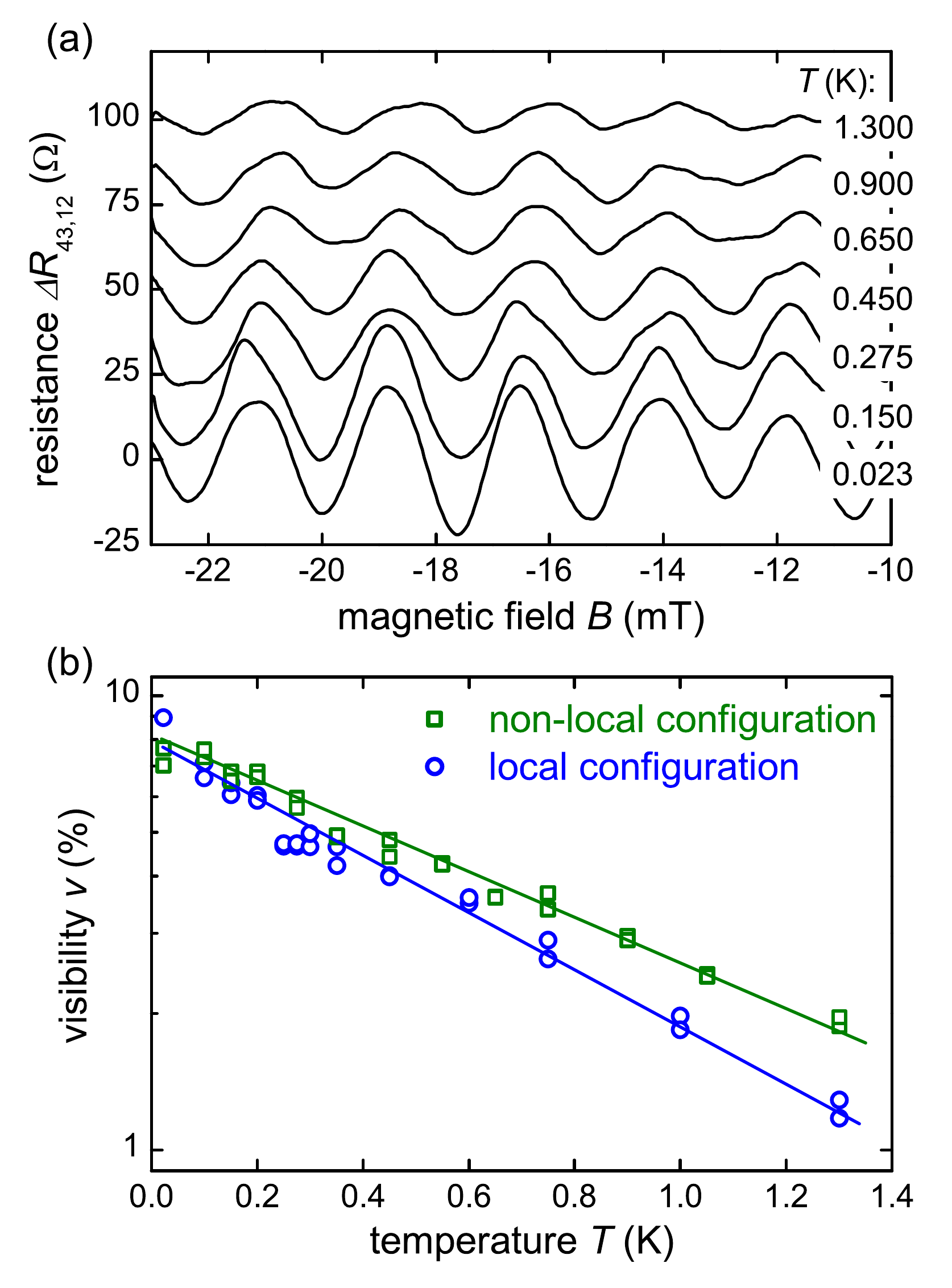}
\caption{(Color online) (a) Oscillatory components of the non-local resistance $R_{43,12}$ (device B, $V_\mathrm{tg}=0.73$~V) at temperatures between $T_\mathrm{base}=23$~mK and $T=1.3$~K. The measurements are offset for clarity. (b) Temperature dependences of the AB oscillation visibilities of local ($R_{14,23}$) and non-local ($R_{43,12}$) measurements of device B.}
\label{tempdec}
\end{center}
\end{figure}

In this section, we investigate electron dephasing with temperature. Multi-mode transport and a maximum in AB resistance oscillation amplitude were ensured by a proper choice of the applied gate voltages of $V_\mathrm{tg}=0.7$~V and 0.73~V in the local and the non-local configuration of device B, respectively. In order to minimize external effects we chose identical instrumental setups with the same currents in both measurement configurations. The oscillatory components of non-local resistance traces are shown in Fig.~\ref{tempdec}~(a) for temperatures between $T_\mathrm{base}=23$~mK and $T=1.3$~K. As temperature increases, the oscillation amplitude decreases. Fig.~\ref{tempdec}~(b) depicts the temperature dependence of the AB oscillation visibility of device B in the local and the non-local probe configurations. For each data point, 8 oscillation periods were recorded to determine the average visibility $v$.
Assuming a temperature dependence of $v(T)=v_0\exp (-\alpha T)$, the dephasing rates $\alpha$ are $\alpha_\mathrm{loc}=1.44$~K$^{-1}$ and $\alpha_\mathrm{n-loc}=1.16$~K$^{-1}$ in the local and the non-local probe configurations, respectively.

The attenuation of the AB oscillation visibility can be approximated \cite{hans01,seel01} as $v\propto\exp(-\tau/\tau_\mathrm{deph})$, where in our device $\tau=s/v_\mathrm{F}$ is the propagation time along the mean path of the ring $s=(s_1+s_2)/2$ and $\tau_\mathrm{deph}^{-1}$ is the total dephasing rate accounting for all dephasing mechanisms. In general, electron dephasing results from electron-phonon, electron-electron and magnetic impurity interactions as well as thermal averaging. In ballistic conductors, especially 1D structures, inter-subband and boundary scattering may come into play as well, and in AB rings elastic scattering randomizes the phase due to a random modification of the electrons' path lengths \cite{aiha91}.

However, at low temperatures the dominating dephasing mechanisms in ballistic conductors are electron-electron scattering \cite{alts82} and thermal averaging \cite{wash85}. Electron-electron scattering leads to phase breaking of the involved electron waves with a rate $\tau_\mathrm{\Phi}^{-1}$ whereas thermal averaging gives rise to phase averaging due to the thermal broadening of roughly $3.5k_\mathrm{B}T$ around the Fermi energy $E_\mathrm{F}$ and can be associated with a rate $\tau_\mathrm{th}^{-1}$. Hence, the total dephasing rate is $\tau_\mathrm{deph}^{-1}=\tau_\mathrm{\Phi}^{-1}+\tau_\mathrm{th}^{-1}$.

The thermal averaging rate $\tau_\mathrm{th}^{-1}$ can be estimated from the phase difference $\Delta k_\mathrm{F}s\approx3.5k_\mathrm{B}T\cdot s/(\hbar v_\mathrm{F})$ between the two electron paths $s_1$ and $s_2$. Here, $\Delta k_\mathrm{F}$ was determined by a first-order approximation of $k_\mathrm{F}$ around $E_\mathrm{F}$. Phase averaging becomes relevant when the propagation time is in the order of the dephasing time $\tau\approx\tau_\mathrm{th}$ and the phase difference is $\Delta k_\mathrm{F}s=\pi$. This yields the thermal dephasing rate $\tau_\mathrm{th}^{-1}\approx k_\mathrm{B}T/\hbar$ and the visibility
\begin{equation}
v=v_0\exp(-\tau/\tau_\mathrm{\Phi})\exp(-sk_\mathrm{B}T/(\hbar v_\mathrm{F})).
\end{equation}
For the device B at a gate voltage of $V_\mathrm{tg}=0.73$~V we estimate the factor $\alpha_\mathrm{th}=sk_\mathrm{B}/(\hbar v_\mathrm{F})\approx1.1$~K$^{-1}$, where $v_\mathrm{F}$ was determined in a 2D-model from Shubnikov-de Haas measurements as mentioned above. Compared with the experimental values ($\alpha_\mathrm{loc}=1.44$~K$^{-1}$ and $\alpha_\mathrm{n-loc}=1.16$~K$^{-1}$), the rough estimate $\alpha_\mathrm{th}\approx1.1$~K$^{-1}$ indicates that thermal averaging is the predominant cause of dephasing in the non-local measurement configuration. 

Recently, Seelig and B\"uttiker proposed a phase-breaking mechanism related to potential fluctuations in the waveguides of the ring being induced by fluctuating voltages at the voltage sensing leads \cite{seel03}. The measurement circuitry can contribute to dephasing with a dephasing rate linear in temperature and depending on the measurement configuration \cite{seel03} as well as gates \cite{seel01}. Significantely different dephasing rates in the local and the non-local measurement configurations are expected if the transmission of electrons from a lead into the two ring arms is asymmetric \cite{seel03}.

Experimentally, the discussed probe-configuration-dependence of the dephasing has been observed in a symmetric AB ring covered by local finger top-gates of 100~nm width \cite{koba02}. The determined  total dephasing rates are around 2.5~K$^{-1}$ and 1~K$^{-1}$ in the local and the non-local probe configurations, respectively, illustrating the drastic influence of the measurement circuitry on decoherence. For our geometrically asymmetric ring with a global top-gate the tendency is the same. Fig.~\ref{tempdec}~(b) shows that dephasing in the local configuration is stronger than in the non-local configuration ($\alpha_\mathrm{loc}=1.44$~K$^{-1}$ and $\alpha_\mathrm{n-loc}=1.16$~K$^{-1}$).

Our findings complement the results on dephasing mechanisms in multi-terminal ballistic 1D AB rings by a globally top-gated interferometer both asymmetric in geometry and electron transmission. The transmission asymmetry, here realized by orthogonal cross-junctions, is expected to increase the measurement-configuration-dependence of dephasing \cite{seel03}.
A full clarification of the observed dephasing will require a theoretical approach which takes into account the specific geometry, the asymmetric transmission at the orthogonal waveguide cross-junctions and the global top-gate.

\section{Conclusion}

In summary, we have investigated asymmetric four-terminal waveguide quantum rings with respect to electron phase sensitivity and thermal dephasing. A global top-gate enables the electrostatic control of the transmission phase determined from AB resistance oscillations.
In the non-local four-probe measurement configuration a continuous transmission phase shift was observed. The magnitude of this shift was roughly approximated with a single-electron model and is in good agreement with the experimental observation. We attribute the general observation of a phase shift to (a) the strongly asymmetric quantum ring geometry breaking transmission symmetries, (b) the realization of individual 1D QWR voltage and current leads, (c) the distance between the 1D leads which is smaller than the phase coherence length and to (d) minimized reflections between current and voltage leads due to orthogonal waveguide cross-junctions. 

The temperature dependence of AB oscillations yields different dephasing rates in the local and the non-local measurement configurations. This demonstrate the influence of the measurement circuitry on dephasing. The AB oscillation amplitude decreases exponentially with temperature. While the non-local measurement is dominated by thermal averaging alone, potential fluctuations are an additional source of decoherence in the local measurement.

Our results demonstrate the control of the transmission phase and the accurate determination of its shift in an electron waveguide AB interferometer. The investigated 1D-geometry appears promising for future studies of the coherence and transmission phase in single-mode transport with respect to correlation and spin effects of embedded or attached quantum devices and circuits.

\section*{Acknowledgments}

We thank E. Sternemann, C. Kreisbeck and T. Kramer for helpful discussions. The authors gratefully acknowledge financial support from the Deutsche Forschungsgemeinschaft within the the priority program SPP1285. SSB and MB gratefully acknowledge support by the Research School of the Ruhr-Universi\"at Bochum. SFF gratefully acknowledges support by the Alexander-von-Humboldt Foundation.


\begin{thebibliography}{10}
\bibitem{ahar59} Y. Aharonov, D. Bohm, Phys. Rev. \textbf{115}, 485 (1959).
\bibitem{yac95} A. Yacoby, M. Heiblum, D. Mahalu, H. Shtrikman, Phys. Rev. Lett. \textbf{74}, 4047 (1995).
%
\bibitem{schu97} R. Schuster, E. Buks, M. Heiblum, D. Mahalu, V. Umansky, H. Shtirkman, Nature \textbf{385}, 417 (1997).
\bibitem{koba0204} K. Kobayashi, H. Aikawa, S. Katsumoto, Y. Iye, Phys. Rev. Lett. \textbf{88}, 256806 (2002); K. Kobayashi, H. Aikawa, A. Sano, S. Katsumoto, Y. Iye, Phys. Rev. B \textbf{70}, 035319 (2004).
\bibitem{sigr04}
M. Sigrist, A. Fuhrer, T. Ihn, K. Ensslin, S.E. Ulloa, W. Wegscheider, M. Bichler, Phys. Rev. Lett. \textbf{93}, 066802 (2004).
\bibitem{ji00} Y. Ji, M. Heiblum, D. Sprinzak, D. Mahalu, H. Shtrikman, Science \textbf{290}, 779 (2000); Y. Ji, M. Heiblum, H. Shtrikman, Phys. Rev. Lett. \textbf{88}, 076601 (2002); M. Avinun-Kalish, M. Heiblum, O. Zarchin, D. Mahalu, V. Umansky, Nature \textbf{436}, 529 (2005); M. Zaffalon, A. Bid, M. Heiblum, D. Mahalu, V. Umansky, Phys. Rev. Lett. \textbf{100}, 226601 (2008).
\bibitem{koen02} J. K\"onig, Y. Gefen, Phys. Rev. B \textbf{65}, 045316 (2002).
\bibitem{koen10} J. K\"onig, B. Hiltscher, private communication (2010).
%
\bibitem{yaco96} A. Yacoby, R. Schuster, M. Heiblum, Phys. Rev. B \textbf{53}, 9583 (1996).
\bibitem{butt88} M. B\"uttiker, IBM J. Res. Develop. \textbf{32}, 317 (1988).
\bibitem{enti02} O. Entin-Wohlman, A. Aharony, Y. Imry, Y. Levinson, A. Schiller, Phys. Rev. Lett. \textbf{88}, 166801 (2002); A. Aharony, O. Entin-Wohlman, B.I. Halperin,, Y. Imry, Phys. Rev. B \textbf{66}, 115311 (2002).
\bibitem{krei10} C. Kreisbeck, T. Kramer, private communication (2010).
\bibitem{datt89} S. Datta, Superlattices Microstruct. \textbf{6}, 83 (1989).
%
\bibitem{koba02} K. Kobayashi, H. Aikawa, S. Katsumoto, Y. Iye, J. Phys. Soc. Jpn. \textbf{71}, 2094 (2002).
\bibitem{aiha91} K. Aihara, M. Yamamoto, K. Iwadate, T. Mizutani, Jap. J. Appl. Phys. \textbf{30}, 1627 (1991).
\bibitem{cass00} M. Casse, Z.D. Kvon, G.M. Gusev, E.B. Olshanetskii, L.V. Litvin, A.V. Plotnikov, D.K. Maude, J.C. Portal, Phys. Rev. B \textbf{62}, 2624 (2000).
\bibitem{hans01} A.E. Hansen, A. Kristensen, S. Pedersen, C.B. Sorensen, P.E. Lindelof, Phys. Rev. B \textbf{64}, 045327 (2001).
\bibitem{yama09} Y. Yamauchi, M. Hashisaka, S. Nakamura, K. Chida, S. Kasai, T. Ono, R. Leturcq, K. Ensslin, D.C. Driscoll, A.C. Gossard, K. Kobayashi, Phys. Rev. B \textbf{79}, 161306(R) (2009).
\bibitem{lin10} K.-T. Lin, Y. Lin, C.C. Chi, J.C. Chen, T. Ueda, S. Komiyama, Phys. Rev. B \textbf{81}, 035312 (2010).
\bibitem{seel03} G. Seelig, S. Pilgram, A.N. Jordan, M. B\"uttiker, Phys. Rev. B \textbf{68}, 161310(R) (2003).
\bibitem{seel01} G. Seelig, M. B\"uttiker, Phys. Rev. B \textbf{64}, 245313 (2001).
\bibitem{alts82} B.L. Altshuler, A.G. Aronov, D.E. Khmelnitzky, J. Phys. C: Sol. State Phys. \textbf{15}, 7367 (1982).
\bibitem{wash85} S. Washburn, C.P. Umbach, R.B. Laibowitz, R.A. Webb, Phys. Rev. B \textbf{32}, 4789 (1985).
\bibitem{buch09a} S.S. Buchholz, S.F. Fischer, U. Kunze, D. Reuter, A.D. Wieck, Appl. Phys. Lett., \textbf{94}, 022107 (2009).
\bibitem{buch09b} S.S. Buchholz, S.F. Fischer, U. Kunze, D. Reuter, A.D. Wieck, Physica E \textbf{42}, 1099 (2009).
\bibitem{vanW88} B.J. van Wees, H. van Houten, C.W.J. Beenakker, J.G. Williamson, L.P. Kouwenhoven, D. van der Marel, C.T. Foxon, Phys. Rev. Lett. \textbf{60}, 848 (1988).
\bibitem{whar88} D.A. Wharam, T.J. Thornton, R. Newbury, M. Pepper, H. Ahmed, J.E.F. Frost, D.G. Hasko, D.C. Peacockt, D.A. Ritchie, G.A.C. Jones, J. Phys. C: Sol. State Phys. \textbf{21}, L209 (1988).
\bibitem{lian99} C.-T. Liang, M.Y. Simmons, C.G. Smith, D.A. Ritchie, M. Pepper, Appl. Phys. Lett. \textbf{57}, 2975 (1999).
\bibitem{knop05} M. Knop, M. Richter, R. Massmann, U. Wieser, U. Kunze,
D. Reuter, C. Riedesel, A.D. Wieck, Semicond. Sci. Technol. \textbf{20}, 814 (2005).
\bibitem{onsa31} L. Onsager, Phys. Rev. \textbf{38}, 2265 (1931); H.B.G. Casimir, Rev. Mod. Phys. \textbf{17}, 343 (1945).
\bibitem{butt86} M. B\"uttiker, Phys. Rev. Lett. \textbf{57}, 1761 (1986).
\bibitem{ped00} S. Pedersen, A.E. Hansen, A. Kristensen, C.B. Sorensen, P.E. Lindelof, Phys. Rev. B \textbf{61}, 5457 (2000).
\bibitem{stra09} E. Strambini, V. Piazza, G. Biasiol, L. Sorba, F. Beltram, Phys. Rev. B \textbf{79}, 195443 (2009).
\bibitem{wash87} S. Washburn, H. Schmid, D. Kern, R.A. Webb, Phys. Rev. Lett. \textbf{59}, 1791 (1987).
\bibitem{deVeg89} P.G.N. de Vegvar, G. Timp, P.M. Mankiewich, R. Behringer, J. Cunningham, Phys. Rev. B \textbf{40}, 3491 (1989).
\bibitem{ford90} C.J.B. Ford, A.B. Fowler, J.M. Hong, C.M. Knoedler, S.E. Laux, J.J. Wainer, S. Washburn, Surf. Sci. \textbf{229}, 307 (1990).
%
\bibitem{kram08} T. Kramer, E.J. Heller, R.E. Parrott, J. Phys. conf. ser. \textbf{99}, 012010 (2008); T. Kramer, C. Kreisbeck, V. Krueckl, arXiv p.1002.5042 (2010).
\bibitem{cern97} G. Cernicchiaro, T. Martin, K. Hasselbach, D. Mailly, A. Benoit, Phys. Rev. Lett. \textbf{79}, 273 (1997).
\bibitem{kraf01} B. Krafft, A. F\"orster, A. van der Hart, T. Sch\"apers, Physica E \textbf{9}, 635 (2001).
\bibitem{ben86} A.D. Benoit, S. Washburn, C.P. Umbach, R.B. Laibowitz, R.A. Webb, Phys. Rev. Lett. \textbf{57}, 1765 (1986).

\end{thebibliography}
\end{document}